\documentclass[preprint,showpacs,showkeys,preprintnumbers,amsmath,amssymb,aps]{revtex4}
\usepackage{epsf}
\usepackage{graphicx}% Include figure files
\usepackage{dcolumn}% Align table columns on decimal point
\usepackage{bm}% bold math
\everymath{\displaystyle}
\begin{document}

\title{Effect of Electric and Magnetic Fields on Spin Dynamics in the Resonant Electric
Dipole Moment Experiment}

\author{Alexander J. Silenko}

\affiliation{Institute of Nuclear Problems, Belarusian State
University, 11 Bobruiskaya Street, Minsk 220080, Belarus} \email{
silenko@inp.minsk.by}

\date{\today}
\begin {abstract} A buildup of the vertical polarization in the resonant
electric dipole moment (EDM) experiment [Y. F. Orlov, W. M. Morse,
and Y. K. Semertzidis, Phys. Rev. Lett. {\bf 96}, 214802 (2006)]
is affected by a horizontal electric field in the particle rest
frame oscillating at a resonant frequency. This field is defined
by the Lorentz transformation of an oscillating longitudinal
electric field and a
uniform vertical magnetic one. %given in the lab frame
The effect of a longitudinal electric field is significant, while
the contribution from a magnetic field caused by forced coherent
longitudinal oscillations of particles is dominant. The effect of
electric field on the spin dynamics was not taken into account in
previous calculations. This effect is considerable and leads to
decreasing the EDM effect for the deuteron and increasing it for
the proton. The formula for resonance strengths in the EDM
experiment has been derived. The spin dynamics has been
calculated.
\end{abstract}
 \maketitle

\section {Introduction}

Sensitive searches for electric dipole moments (EDMs) of
fundamental particles are excellent probes of physics beyond the
Standard Model \cite{EDMphys}. The Storage Ring EDM Collaboration
investigates the possibility to search for a particle EDM by
placing a charged particle in a storage ring and observing the
precession of its spin \cite{EDM,EDMP}. The frozen spin method
\cite{EDM,EDMP,EDMNSPC} which consists in cancelling a spin
precession in the horizontal plane with a radial electric field
can provide a sensitivity of deuteron EDM measurements of
$10^{-27}$ e$\cdot$cm \cite{EDMP}.

The resonance method proposed by Y. Orlov \cite{YOr} reaches the
sensitivity of $10^{-29}$ e$\cdot$cm for the deuteron and
$10^{-28}$ e$\cdot$cm for the proton. This method is based on the
idea that the precession caused by the deuteron EDM will
accumulate if the beam is forced to undergo a coherent
longitudinal oscillation that is in phase with the spin precession
in the horizontal plane. This oscillation is caused by an
oscillating longitudinal electric field and its frequency should
be very close to that of the spin precession. The resonance effect
in the storage ring (so-called Orlov ring) is ensured by radio
frequency (rf) cavities \cite{YOr,YS,OMS}.

The resonance method is free of main systematic error of
measurement of the EDM with the frozen spin method caused by the
presence of a small vertical electric field (see Refs.
\cite{EDM,EDMP}). This field leads to
the buildup of the vertical polarization (BVP) %of particles
imitating the EDM effect \cite{EDM,EDMP}.

The aim of this work is the calculation of effects of electric and
magnetic fields on a buildup of the vertical polarization in a
resonant EDM experiment \cite{YOr,YS,OMS}. Formulae for resonance
strengths are derived. Distinguishing features of the resonant EDM
experiment for protons and deuterons are considered and dynamics
of components of polarization vector is described.

Throughout the work the system of units $\hbar=c=1$ is used. In
some equations, the speed of light $c$ is explicitly shown.

\section {Fields Affecting the Buildup of the Vertical Polarization}

In the resonance EDM experiment \cite{YOr,YS,OMS}, it is planned
to stimulate the BVP conditioned by the EDM and to avoid a similar
effect caused by the magnetic moment. It is known that the
magnetic resonance takes place when a particle placed in a uniform
vertical magnetic field is also affected by a horizontal magnetic
field oscillating at a frequency close to the frequency of spin
rotation (see, e.g., Ref. \cite{Slic}). If the particle moves, the
magnetic resonance can also be stimulated by an oscillating
electric field transforming to an oscillating magnetic one in the
particle rest frame. The magnetic resonance results in spin
flipping for a vertically polarized beam and in the BVP for a
horizontally polarized one.

In this work, ``spin" means an expectation value of a quantum
mechanical spin operator. The polarization vector is defined by
$\bm P=\bm S/S$, where $\bm S$ is the spin vector and $S$ is the spin quantum
number. The directions of the EDM vector, $\bm d$, and the spin coincide:
$\bm d=d\bm S/S$. It is convenient to use the $\eta$ factor for the EDM which corresponds
to the $g$ factor for the magnetic moment and is given by
$$\eta=\frac{2dm}{eS}.$$

Evidently, the magnetic resonance cannot take place when the
electric field is longitudinal, because nothing but the
oscillating electric field appears in the particle rest frame.
Since the frequencies of betatron oscillations are chosen to be
far from resonances, these oscillations cannot lead to the
resonance effect. However, the resonance takes place if the
particle possesses the EDM. The resonance is ``electric", because it is conditioned by
the electric field in the particle rest frame. In this frame, the
electric field possesses the longitudinal component $E'_\phi$
defined by the oscillating electric field and the radial component
$E'_\rho$ caused by the Lorentz transformation of the vertical
magnetic field. The latter component has a resonance part because
of the modulation of the particle velocity. Only this component
has been taken into account in previous calculations
\cite{YOr,YS,OMS}. The resonance effect is provided by both
components of the electric field in the particle rest frame.

We use the particle rest frame for explaining the origin of the
resonance. However, we do not utilize it for calculations and
derive all basic equations in a cylindrical coordinate system. The
particle spin motion in storage rings is usually specified with
respect to the particle trajectory. Main fields are commonly
defined relative to the cylindrical coordinate axes. When the ring
is either circular or divided into circular sectors by empty
spaces, the use of cylindrical coordinates considerably simplifies
the analysis of spin effects \cite{N84}. The equation of spin
motion in the cylindrical coordinate system coincides with that in
the frame rotating together with the particle (rotating frame)
because the horizontal axes of the cylindrical coordinate system
rotate at the instantaneous angular frequency of orbital
revolution. The motion of particles in the rotating frame is
relatively slow because it can be caused only by %beam
oscillations and other deflections of the particles from the ideal
trajectory. Therefore, the difference between the spin motion in
the cylindrical coordinate system and the particle rest frame can
be neglected in many cases.

The general equation of spin motion in the cylindrical coordinates
is given by \cite{N84}
\begin{eqnarray} \frac{d\bm S}{dt}=\bm\omega_a\times\bm S,
\nonumber\\ \bm\omega_a=-\frac{e}{m}\left\{a\bm B-
\frac{a\gamma}{\gamma+1}\bm\beta(\bm\beta\cdot\bm B)\right.\nonumber\\
+\left(\frac{1}{\gamma^2-1}-a\right)\left(\bm\beta\times\bm
E\right)%\nonumber\\
+\frac{1}{\gamma}\left[\bm B_\|
-\frac{1}{\beta^2}\left(\bm\beta\times\bm
E\right)_\|\right]\nonumber\\
\left.+ \frac{\eta}{2}\left(\bm
E-\frac{\gamma}{\gamma+1}\bm\beta(\bm\beta\cdot\bm
E)+\bm\beta\!\times\!\bm B\right)\!\right\},\nonumber\\
\bm\beta=\frac{\bm v}{c},~~~ a=\frac{g-2}{2},
\label{eq8}\end{eqnarray} where $\bm\omega_a$
is the angular velocity of spin precession relative to axes of the
cylindrical coordinate system.
%In this equation, the fields are given in the lab frame.

The sign $\|$ means a horizontal
projection for any vector. Fields defining perturbations of particle
trajectory significantly affect the spin motion, while the
explicit dependence of the quantity $\bm\omega_a$ on radial and
vertical components of particle momentum conditioning these
perturbations can usually be neglected (see Ref. \cite{N84}).

To measure the effect, some resonators (rf cavities) should be
used. The electric field in a resonator is generated along the
central line, and the magnetic field is orthogonally directed
\cite{JJ}. The magnetic field along the central line is equal to
zero. If the rf cavities are perfectly placed and longitudinally
directed, the magnetic field cannot stimulate any resonance
effect. The resonant effect leading to the regular BVP is caused
by the terms proportional to $\eta$. Therefore, the observed BVP
corresponds to the definite value of the EDM. However, both a
displacement and an angular deviation of the center line of the rf
cavities away from an average particle trajectory lead to a
similar behavior of spin imitating the EDM effect. As a result,
they create systematic errors in the measurement of the EDM. Most
of these errors are not in resonance with the spin precession in
the horizontal plane. Therefore, they create background and result
in fast oscillations of the vertical component of the polarization
vector \cite{YOr,YS,OMS,YSNotes}. Besides this effect, the
systematic error can be caused by a radial magnetic field in the
particle rest frame oscillating at the resonant frequency. This
error will be eliminated by alternately producing two sub-beams
with different betatron tunes \cite{YOr,OMS,YSNotes}. In the
present work, we calculate only the effects of resonant fields on
the BVP in ideal conditions and disregard systematic errors.

Since the velocity oscillates, the vertical magnetic field creates
the resonance part of the radial electric field in the particle
rest frame. Thus, we need to take into consideration the constant
vertical magnetic field and the oscillating longitudinal electric
one in the lab frame. This is properly shown by Eq. (\ref{eq8}).
Resonant terms in the expression for the angular velocity of spin
rotation are proportional to the EDM:
\begin{equation}  \bm\Omega_{EDM}=-\frac{e\eta}{2m}\left[\bm
E-\frac{\gamma}{\gamma+1}\bm\beta(\bm\beta\cdot\bm
E)+\bm\beta\!\times\!\bm B\right]. \label{eq2}
\end{equation}
Eq. (\ref{eq2}) describes the interaction of the EDM with the
electric field in the rotating frame in terms of the lab frame
fields.

The polarization of the circulating deuteron beam can be best
measured when colliding the deuterons with a carbon target and
observing reaction products generated due to nuclear interactions
\cite{EDMP,COSY}. The polarization of the circulating proton beam
can be determined by means of elastic proton-proton scattering
(see Refs. \cite{COSY,EDDA} and references therein).

\section {Resonance Strengths in the EDM Experiment}

%In the deuteron EDM experiment,
The longitudinal electric field acting
on the deuterons in the resonator oscillates at the angular frequency $\omega$ which
%is equal to the difference between two
%radio frequencies. It
should be very close to the angular frequency of spin rotation
(g$-$2 frequency), $\omega_0$, and %close
to the eigenfrequency of
free synchrotron oscillations (synchrotron frequency) \cite{OMS}.
The quantity $\omega_0$ is almost equal to the vertical component
of $\bm\omega_a$, because other components of this (pseudo)vector
are relatively small:
\begin{equation} \omega_0=
\left(\omega_a\right)_z=-\frac{ea}{m}B_0.
\label{eqom}\end{equation} In Eq. (\ref{eqom}), $B_0$ is the
average vertical magnetic field. We suppose the particle charge to
be positive and the magnetic field to be upward ($B_0>0$). The
cyclotron frequency is given by
\begin{equation} \omega_c=-\frac{eB_0}{\gamma_0 m},
\label{frequency}\end{equation} where $\gamma_0$ is the average
Lorentz factor. The minus sign means that the particle rotates
clockwise ($\bm\beta\cdot\bm e_\phi<0,~\omega_c<0$).
%where [$\left(\omega_c\right)_z=\bm\omega_c\cdot\bm e_z<0$].

Because the oscillating electric field is longitudinally directed,
\begin{equation}  \bm E-\frac{\gamma}{\gamma+1}\bm\beta(\bm\beta\cdot\bm E)=
\frac{1}{\gamma}\bm E. \label{eq4} \end{equation}

If we take into account only fields conditioning the electric
field in the particle rest frame, the spin motion in the
horizontal plane is defined by Eq. (\ref{eq2}), where
$\bm\beta=-\beta\bm e_\phi$.

The action of resonant electric and quasi-electric ($\bm
G\equiv\bm\beta\times\bm B$) fields on the EDM is similar to that
of resonant magnetic and quasi-magnetic ($-\bm\beta\times\bm E$)
fields on the magnetic moment. Therefore, some previously obtained
expressions for resonance strengths (see Refs.
\cite{Roser,Bai,LeePRSTAB} and references therein) may be utilized
when analyzing EDM-dependent interactions. For the EDM experiment,
two regimes with large coherent oscillations are possible: a
strongly linear regime (using, for example, a specially designed
rf cavity for the linearization of oscillations), or a strongly
nonlinear regime with well-stabilized coherent oscillations (see
Ref. \cite{OMS} and references therein). We confine ourselves to
the consideration of linear oscillations.

It is convenient to use the quantity
\begin{eqnarray} \Phi=\phi(t)-\phi(0)=\omega_c t,
\label{Phi}\end{eqnarray} where $\phi$ is the azimuth
of the particle at the given moment of time. The distance
traversed by the beam is equal to
$$L_b=\frac{\beta c}{\omega_c}\Phi=\beta ct.$$

Similarly to the magnetic fields of a rf dipole and a rf solenoid
\cite{Roser,Bai,LeePRSTAB}, the longitudinal electric field of a
rf cavity can be expressed in terms of delta functions:
\begin{eqnarray} \bm E=E_0\frac{L(\Phi)}{\rho}\sin{(\omega t+\varphi)}\bm
e_\phi, ~~~ %\nonumber\\
L(\Phi)=l\sum^\infty_{N=-\infty}{\delta(\Phi-\Phi_0-2\pi N)}, ~~~
%\nonumber\\
\rho=-\frac{\beta c}{\omega_c}, \label{delta}\end{eqnarray} where
$E_0,\omega$ and $\varphi$ are the amplitude, angular frequency
and phase of electric field acting on the particle in the resonator,
$l$ and $\Phi_0$ define the length and position (azimuth) of the resonator,
and $\rho$ is the radius of
curvature. %The quantity $\bm E$ can be defined in the lab frame at
%a given point of particle trajectory.
%Eq. (\ref{eq2}) describes the interaction of the EDM with the
%electric field in the rotating frame in terms of the lab frame
%fields.
%To provide a significant synchrotron oscillation
%of the beam at the resonance frequency, the angular frequency of the
%resonator $\omega$ should be very close to the eigenfrequency of
%free synchrotron oscillations (synchrotron frequency) \cite{OMS}.
Eqs. (\ref{Phi}),(\ref{delta}) result in
\begin{eqnarray} \bm E=\frac{E_0l}{\rho}\sin{(\nu\Phi+\varphi)}
\sum^\infty_{N=-\infty}{\delta(\Phi-\Phi_0-2\pi N)}\bm
e_\phi, \label{delti}\end{eqnarray} where $\nu=\omega/\omega_c$.

The equation of particle motion is
given by
%The particle momentum is defined by means of integration of Eq.
\begin{equation}
\frac{d\bm p}{dt}=e\bm E. \label{eqo} \end{equation}
%defines the particle momentum.
To determine the particle momentum, one needs to integrate this equation.
The integral of the delta function is the Heaviside function:
$$\int_{-\infty}^\Phi{\delta(\phi-\Phi_0-2\pi N)d\phi}=H(\Phi-\Phi_0-2\pi N),
~~~H(x)=\left\{\begin{array}{c} 0,~~~~~x<0 \\ 1/2,~~~x=0\\ 1,~~~~~x>0 \end{array}\right..$$
Therefore, the momentum is described by the piecewise continuous function:
%being the series of Heaviside functions:
%taking the form
\begin{eqnarray} \bm p=\frac{eE_0l}{\rho}
\sum^\infty_{N=-\infty}{\sin{(\nu\Phi_N+\varphi)}H(\Phi-\Phi_N)}\bm
e_\phi+\bm C, \label{eqHf}\end{eqnarray} where $\Phi_N=\Phi_0+2\pi N$ and
$\bm C=const$. The momentum is either minimum or maximum when $\sin{(\nu\Phi_N+\varphi)}
\approx0$. Its change defined by the height of a step is
maximum when $\sin{(\nu\Phi_N+\varphi)}\approx\pm1$. Therefore, the
momentum is modulated by the cosine function $\cos{(\nu\Phi+\varphi)}$.

Spin dynamics described by Eqs. (\ref{eq8}),(\ref{eq2}) is determined by
integration of angular velocity of spin motion. The evolution of vertical
component of the spin is very slow and resonant fields affecting the spin
in the particle rest frame remain almost constant during one revolution.

It would be very inconvenient to use simultaneously the delta and
Heaviside functions. It is better to utilize the fact that function (\ref{eqHf}) can be replaced by another that
integration brings the same result. Therefore, an appropriate delta function can
be substituted for the sum of the Heaviside functions at every step of function (\ref{eqHf}).
As a result, one obtains a series of delta functions modulated by $\cos{(\nu\Phi+\varphi)}$.
The quantity
\begin{eqnarray} \bm\pi=-p_0\bm e_\phi+a\cos{(\nu\Phi+\varphi)}
\sum^\infty_{N=-\infty}{\delta(\Phi-\Phi_0-2\pi N)}\bm
e_\phi \label{eqpf}\end{eqnarray}
may be therefore substituted for the momentum $\bm p$ because the integration of $\bm\pi$ and $\bm p$
leads to the same result. In Eq. (\ref{eqpf}), $\bm p_0$ is the average momentum and $a$ is the constant
that value is defined by the equation of particle motion:
\begin{equation}
\frac{d\bm\pi}{dt}=\omega_c\frac{d\bm\pi}{d\Phi}=e\bm E. \label{eqpi} \end{equation}

Eqs. (\ref{delta}),(\ref{eqpf}),(\ref{eqpi}) result in
\begin{equation} \bm\pi=-\left[p_0+\Delta p_0\cos{(\nu\Phi+\varphi)}\right]\bm e_\phi,~~~ \Delta
p_0=\frac{eE_0L(\Phi)}{\omega\rho}. \label{eqnew}\end{equation}

%Eq. (\ref{eqnew}) gives an opportunity to determine the piece-wise
%constant momentum in the arcs of the ring.

We can perform calculations to within first-order terms in
$\Delta\beta_0$ and neglect beam oscillations at nonresonance
frequencies. The normalized velocity is given by
$$\bm \beta=\frac{\bm p}{\sqrt{m^2+
\bm p^2}}.$$

In this equation, substitution of $\bm\pi$ for $\bm p$ does not change the result of integration of Eq. (\ref{eq2}). Therefore,
the normalized velocity can be replaced by the quantity
\begin{equation} \bm \beta'=\frac{\bm \pi}{\sqrt{m^2+
\bm\pi^2}}=-\left[\beta_0+\Delta\beta'_0\cos{(\nu\Phi+\varphi)}\right]\bm e_\phi, \label{eqvel}\end{equation} where
\begin{equation} \beta_0=\frac{p_0}{m\gamma_0},~~~\gamma_0=\frac{\sqrt{m^2+p_0^2}}{m}, ~~~
\Delta\beta'_0=\frac{\Delta
p_0}{m\gamma_0^3}=\frac{eE_0L(\Phi)}{m\gamma_0^3\omega\rho}.
\label{eqdel}\end{equation}
The primes will be omitted below.

Eqs. (\ref{eq8})--(\ref{eqdel}) result in
\begin{eqnarray} \frac{d\bm S}{d\Phi}=\bm F\times\bm S,
~~~ %\nonumber\\
\bm F=F_1\bm e_\rho+F_2\bm e_\phi+F_3\bm e_z
\nonumber\\=\frac{e\eta}{2p_0}E_0L(\Phi)\Bigl[\frac{\omega_0}
{a\gamma_0^2\omega}\cos{(\nu\Phi+\varphi)}\bm e_\rho %\nonumber\\
%\right.\\\left.
+\sin{(\nu\Phi+\varphi)}\bm e_\phi\Bigr]+a\gamma_0\bm e_z,
\label{eq8n}\end{eqnarray} where $F_3=a\gamma_0$ is the spin
precession tune.
%%%%%%%%%%%%%%%%%%%%%%%%%%%%%%%%%%%%%%%%%%%%%%%%% and $\nu=\omega/\omega_c$.

It is convenient to use the method
%rewrite Eq. (\ref{eq8n}) in terms
of resonant strengths
%\begin{equation}  F_1=-\frac{1}{2}\eta\beta\gamma_0, ~~~
%F_2=-\frac{e\eta}{2m\gamma\omega_c}E_\phi. \label{eqnef}
%\end{equation}
based on the Fourier-series expansion (see Refs.
\cite{CR,Lee,MZ}):
\begin{equation}  F_1-iF_2=\sum^\infty_{K=-\infty}{\epsilon_K e^{-i\nu_K\Phi}}, ~~~
\nu_K=\omega_K/\omega_c. \label{resstrn} \end{equation} In this
equation, the resonance strengths $\epsilon_K$ are Fourier
amplitudes corresponding to spin resonance tunes $\nu_K$ and
$\omega_K$ are harmonic frequencies. In the considered case
\cite{CR,Roser}
\begin{equation}
\nu_K=K\pm\nu, ~~~ K=0,\pm1,\pm2,\dots \label{int}
\end{equation}

The resonance strengths are defined by
\begin{equation}
\epsilon_K=\frac{1}{2\pi
N}\int_{(N)}{(F_1-iF_2)e^{i\nu_K\Phi}d\Phi}, \label{epsk}
\end{equation}
where an integration over the infinite number of turns
$N\rightarrow\infty$ should be carried out.

Eqs. (\ref{delta}),(\ref{eq8n}),(\ref{int}),(\ref{epsk}) lead to
the following expressions for the resonance strengths:
\begin{eqnarray}
\epsilon^+_K=\frac{e\eta}{8\pi p_0}E_0l\left(\frac{\omega_0}
{a\gamma_0^2\omega}+1\right)e^{i(K\Phi_0-\varphi)},
\nonumber\\
\epsilon^-_K=\frac{e\eta}{8\pi p_0}E_0l\left(\frac{\omega_0}
{a\gamma_0^2\omega}-1\right)e^{i(K\Phi_0+\varphi)}, \label{eqres}
\end{eqnarray} where $\epsilon^+_K$ and $\epsilon^-_K$ correspond
to the plus and minus signs in Eq. (\ref{int}), respectively.
Certainly, the dependence of the exponentials from $K$ can be
eliminated by appropriate choice of the initial phase
($\Phi_0=0$).

Eq. (\ref{eqres}) can also be deduced by means of the Fourier
expansion of delta functions (see Ref. \cite{LeePRSTAB}):
\begin{eqnarray} \sin{(\nu\Phi+\varphi)}\sum^\infty_{N=-\infty}{\delta(\Phi-\Phi_0-2\pi N)}
=\frac{1}{2\pi}\sum^\infty_{N=-\infty}{\sin{[\pm
N(\Phi-\Phi_0)+\nu\Phi+\varphi]}},
\nonumber\\
\cos{(\nu\Phi+\varphi)}\sum^\infty_{N=-\infty}{\delta(\Phi-\Phi_0-2\pi
N)} =\frac{1}{2\pi}\sum^\infty_{N=-\infty}{\cos{[\pm
N(\Phi-\Phi_0)+\nu\Phi+\varphi]}}. \label{eqf}\end{eqnarray}

Eqs. (\ref{eq8n}) and (\ref{eqres}) define the relationship
between the quantities $F_1,F_2$ and the resonance strengths which
agrees with similar relationships derived in Refs.
\cite{Roser,Bai,LeePRSTAB} for localized rf magnetic fields. It is
of importance because the results obtained in Refs.
\cite{Roser,Bai} have been called in question
\cite{Morozov,MorozovR,Leonova}.

The main difference between Eq. (\ref{eqres}) and the
corresponding equations for the resonance strengths deduced in
Refs. \cite{Roser,Bai,LeePRSTAB} consists in the noncoincidence of
expressions for $\epsilon^+_K$ and $\epsilon^-_K$. The
noncoincidence is caused by a more complicated form of the spin
precession vector
$\bm F$ in the investigated case. % of the EDM experiment.
Eq. (\ref{eq8n}) shows that this vector possesses nonzero
projections onto two horizontal axes dephased by $\pi/2$.

The substitution of $\epsilon_K$ into Eq. (\ref{resstrn}) results
in
\begin{eqnarray}
Q=F_1-iF_2=\sum^\infty_{K=-\infty}{\left[\epsilon^+_Ke^{-i(K+\nu)\Phi}+
\epsilon^-_Ke^{-i(K-\nu)\Phi}\right]}
\nonumber\\=\frac{e\eta}{8\pi
p_0}E_0l\sum^\infty_{K=-\infty}{\biggl\{\left(\frac{\omega_0}
{a\gamma_0^2\omega}+1\right)e^{-i[K(\Phi-\Phi_0)+\nu\Phi+\varphi]}}
\nonumber\\
+ \left(\frac{\omega_0}
{a\gamma_0^2\omega}-1\right)e^{-i[K(\Phi-\Phi_0)-\nu\Phi-\varphi]}
\biggr\}. \label{exp}\end{eqnarray}

Since $F_1$ and $F_2$ are real, they are given by
\begin{eqnarray}
F_1={\rm Re}(Q), ~~~ F_2=-{\rm Im}(Q). \label{expri}\end{eqnarray}
As a result, Eq. (\ref{eq8n}) takes the form
\begin{eqnarray} \frac{d\bm S}{d\Phi}=[{\rm Re}(Q)\bm e_\rho
-{\rm Im}(Q)\bm e_\phi+a\gamma_0\bm e_z]\times\bm S \nonumber\\
=\Biggl(\sum^\infty_{K=-\infty}{\biggl\{{\rm Re}\left[\epsilon^+_K
e^{-i(K+\nu)\Phi}+\epsilon^-_K e^{-i(K-\nu)\Phi}\right]\bm e_\rho}
\nonumber\\
-{\rm Im}\left[\epsilon^+_K e^{-i(K+\nu)\Phi}+\epsilon^-_K
e^{-i(K-\nu)\Phi}\right]\bm e_\phi\biggr\}+a\gamma_0\bm
e_z\Biggl)\times\bm S. \label{eq8f}\end{eqnarray}

Eqs. (\ref{eqres}) and (\ref{eq8f}) completely define the spin
dynamics in the EDM experiment.

The velocity modulation given by Eqs. (\ref{eqvel}),(\ref{eqdel})
can be expressed in terms of the resonance strengths. The use of
Fourier expansion (\ref{eqf}) brings Eq. (\ref{eqvel}) to the form
\begin{eqnarray} \bm \beta=-\left\{\beta_0+\Delta\beta_m\sum^\infty_{K=-\infty}{\cos{[\pm
K(\Phi-\Phi_0)+\nu\Phi+\varphi]}}\right\}\bm e_\phi,
\label{velm}\end{eqnarray} where the amplitude of the velocity
modulation is given by
\begin{eqnarray} \Delta\beta_m=-\frac{e\omega_0}{2\pi a\gamma_0^3
p_0\omega}E_0l. \label{velma}\end{eqnarray}

Eqs. (\ref{eqres}),(\ref{velm}),(\ref{velma}) result in
\begin{eqnarray} \bm \beta=\left\{-\beta_0+\frac{2}{\eta\gamma_0}\left[
\left(1+\frac{a\gamma_0^2\omega}{\omega_0}\right)^{-1}\sum^\infty_{K=-\infty}{\epsilon^+_K
e^{-i(K+\nu)\Phi}} \right.\right.\nonumber\\ \left.\left.
+\left(1-\frac{a\gamma_0^2\omega}{\omega_0}\right)^{-1}
\sum^\infty_{K=-\infty}{\epsilon^-_Ke^{-i(K-\nu)\Phi}}\right]\right\}\bm
e_\phi. \label{velf}\end{eqnarray}

The relation between the resonance strengths and the amplitude of
the velocity modulation has the form
\begin{eqnarray}
\epsilon^+_K=-\frac{\eta\gamma_0}{4}
\left(1+\frac{a\gamma_0^2\omega}{\omega_0}\right)\Delta\beta_m
e^{i(K\Phi_0-\varphi)},
\nonumber\\
\epsilon^-_K=-\frac{\eta\gamma_0}{4}
\left(1-\frac{a\gamma_0^2\omega}{\omega_0}\right)\Delta\beta_m
e^{i(K\Phi_0+\varphi)}. \label{revel}\end{eqnarray}

\section {Resonance Effects of Electric and Magnetic Fields}

The spin rotates counterclockwise (clockwise) if $\omega_0$ is
positive (negative). The resonance field should rotate in the same
direction and its frequency should be very close to the spin
rotation frequency ($\omega_0\approx\omega_c\nu_R$ and
$a\gamma_0\approx\nu_R$, where the index $K=R$ defines the
resonance). The effect of nonresonance harmonics and
backward-rotating fields on the spin motion vanishes on the
average \cite{Slic,CR}. Therefore, Eq. (\ref{eq8f}) contains
nonresonant terms which can be excluded. The resulting equation
takes the form
\begin{eqnarray} \frac{d\bm S}{d\Phi}=\bm F\times\bm S, ~~~ %\nonumber\\
\bm F={\rm Re}\left[\epsilon^\pm_R e^{-i(R\pm\nu)\Phi}\right]\bm
e_\rho -{\rm Im}\left[\epsilon^\pm_R e^{-i(R\pm\nu)\Phi}\right]\bm
e_\phi+a\gamma_0\bm e_z \nonumber\\
=\frac{e\eta}{8\pi
p_0}E_0l\left(\frac{\omega_0}{a\gamma_0^2\omega}\pm1\right)\bm
e_\|+a\gamma_0\bm e_z, \nonumber\\
\bm e_\|=\cos{(\Psi)}\bm e_\rho+\sin{(\Psi)}\bm e_\phi, ~~~
\Psi=R(\Phi-\Phi_0)\pm(\nu\Phi+\varphi),
\label{eqrf}\end{eqnarray} where $\bm e_\|$ is the unit vector
rotating in the horizontal plane with the tune $\nu_R=R\pm\nu$.

The first and second terms in the factor
$\left(\frac{\omega_0}{a\gamma_0^2\omega}\pm1\right)$ are
conditioned by the magnetic and electric fields, respectively. The
relative importance of electric field can be characterized by the
ratio of contributions from the electric and magnetic fields:
\begin{equation}k=\frac{a\gamma_0^2\omega}{\omega_0}. %\approx a\gamma_0^2.
\label{eqr}
\end{equation}
The amplitude of resonant quasi-electric field is given by
\begin{equation}G_0=\frac{E_0}{k\gamma_0}.\label{eqfin}
\end{equation}

Eqs. (\ref{eqrf})--(\ref{eqfin}) are valid for any relation
between $\omega_0$ and $\omega$.

Ratio (\ref{eqr}) is very different for the proton and deuteron.
The gyromagnetic anomaly of the proton is about 12.5 times larger
than that of the deuteron. In addition, the deuteron's
gyromagnetic anomaly is negative. %Therefore, increasing a deuteron
%momentum leads to decreasing the effective rotational field.
The resonance at the frequency
$\omega\approx\omega_0=a\gamma_0\omega_c$ is the dominant harmonic
for the deuteron. For this harmonic, $R=0$. In the planned
deuteron EDM experiment, the quantity $k$ is
equal to $k_d=-0.234$ %$k_d=-0.2344$
for deuterons with momentum
$p_0=1.5$ GeV/c ($\gamma_0=1.28$). %($\gamma_0=1.2805$).
The electric-field correction is significant because it results in
decreasing the EDM effect for the deuteron by %about
23 percent. Although the contribution from the magnetic field to
the BVP is dominant, taking into account this correction is
necessary.
%It should be included in any estimates.

For the proton, both $a_p=1.7928$ and $a_p\gamma_0$ are close to 2
and the angular frequency %of coherent longitudinal oscillations
$\omega=\omega_0$ corresponding to the harmonic $R=0$ is too high
\cite{ES}. Therefore, it is useful to modulate the proton velocity
at the different angular frequency (see Ref. \cite{ES})
$\omega\approx\omega_0-2\omega_c$ which conforms to the spin
resonance tune $\nu_R=2+\nu$ ($\nu$ is positive and $\omega$ is
negative). %This equation can be transformed to the form
%\begin{equation}\omega\approx\omega_{res}=\left(1-\frac{2}{a\gamma_0}\right)\omega_0.\label{eqrp}
%\end{equation}

A possible choice of proton's kinetic energy is $T=161$ MeV
\cite{ES}. This choice gives
$\gamma_0=1.172,~\nu=0.1005,~\omega/\omega_0=\nu/(a\gamma_0)=0.0478$.
%\omega_{res}= 0.0478, \gamma_0=1.1715920
Ratio (\ref{eqr}) is equal to $k_p=0.118$.
%$k_p=0.1177$.
Taking into account the electric-field correction leads to
increasing the EDM effect for the proton by 12 percent. Although
the contribution from the magnetic field is dominant, this
correction is rather important. It should be included in any
estimates.

\section {Resonant Spin Dynamics}

To calculate the resonant spin dynamics, the theory of magnetic
resonance (see Ref. \cite{Slic}) can be used. The spin is governed
by the vertical magnetic field which rotates it in the horizontal
plane and the resonant electric and quasi-electric fields.
%$$\bm E_{eff}=E_{eff}\bm e_\|(t)=\frac{1}{2}B_0\Delta\beta_0
%\left(1+\frac{a\gamma_0^2\omega}{\omega_0}\right).$$
Other fields with spin resonance tunes $\nu_K$ that are far from
the spin rotation tune $a\gamma_0$ can be disregarded. The
resonant spin dynamics is defined by Eq. (\ref{eqrf}) and the
vector $\bm e_\|$ rotates with the tune $\nu_R\approx a\gamma_0$.

To describe the spin dynamics, it is convenient to use the frame
accompanying the spin and rotating at the angular frequency
$\omega_R=\omega_c\nu_R$ (see Ref. \cite{Slic}) relatively to the
cylindrical coordinate axes. The vertical component of the angular
velocity of spin rotation is much less in this frame than in the
cylindrical coordinate system. In the lab frame, the vector $\bm
e_\|$ rotates at an angular frequency which significantly differs
from $\omega$ because of the rotation of the cylindrical
coordinate axes in that frame.

If radial and longitudinal directions in the frame accompanying
the spin coincide with those in the cylindrical coordinate system
at zero time $t=0$, they are defined by the unit vectors
\begin{equation}\begin{array}{c}
\bm e'_\rho=\cos{(\nu_R\Phi)}\bm
e_\rho+\sin{(\nu_R\Phi)}\bm e_\phi, \\%~~~
\bm e'_\phi=-\sin{(\nu_R\Phi)}\bm e_\rho+\cos{(\nu_R\Phi)}\bm
e_\phi.\end{array}\label{eqdob}
\end{equation}
All quantities in the frame accompanying the spin are primed. The angular velocity of spin
rotation is given by \cite{Slic}
\begin{equation}\begin{array}{c}\bm\Omega'=\bm\omega_a-\omega_R\bm
e_z=\omega_c\bm F-\omega_R\bm
e_z,
\end{array}\label{eqt}
\end{equation}
where $\bm F$ is defined by Eq. (\ref{eqrf}). The direction of
vector $\bm\Omega'$ is fixed \cite{Slic} because the unit vector
$\bm e_\|$ transforms to the form
$$\bm e_\|'=\cos{(\zeta')}\bm e'_\rho+\sin{(\zeta')}\bm e'_\phi, ~~~ \zeta'=-R\Phi_0\pm\varphi.$$
%The direction of this vector is defined by the azimuth $\zeta'=-R\Phi_0+\varphi$,
%where $\varphi=0$ corresponds to the radial direction in the lab frame.

The spin rotation tune in the frame accompanying the spin is equal to
\begin{equation}\begin{array}{c}
\nu'=\left|\frac{\bm\Omega'}{\omega_c}\right|=\sqrt{(a\gamma_0-\nu_R^\pm)^2+|\epsilon^\pm_R|^2},
\end{array}\label{eqft}
\end{equation}
where
$$\nu^+_R=R+\nu,~~~ \nu^-_R=R-\nu.$$

The initial beam polarization is supposed to be horizontal in the
resonant EDM ex\-pe\-ri\-ment. The particle spin dynamics depends
on the direction of spin at zero time defined by the azimuth
$\psi$. The azimuth $\psi=0$ corresponds to the initial radial
polarization in the lab frame. The BVP is characterized by the
$z$-component of polarization vector.

Since trajectories of particles in the beam %slightly
depend on their momenta, corresponding values of $\omega_0$ would
vary. However, some experimental techniques will keep the
frequency and phase of the forced coherent longitudinal
oscillations almost equal to the frequency and phase of the spin
rotation. The length of the straight sections is chosen such that
the momentum compaction factor
$$\alpha_p=\frac{\Delta p/p}{\Delta {\cal L}/{\cal L}}=1,$$ where $\Delta
p=p-p_0,~ \Delta {\cal L}={\cal L}-{\cal L}_0,~{\cal L}\equiv
{\cal L}(p),~{\cal L}_0\equiv {\cal L}(p_0)$, and ${\cal L}(p)$ is
the length of the closed orbit for momentum $p$. In this case
$p/{\cal L}=p_0/{\cal L}_0$. Since
$p=m\gamma(p)\omega_c(p)\rho(p)$ and $\rho/{\cal L}=\rho_0/{\cal
L}_0$, the g$-$2 frequency does not depend on the particle
momentum: $a\gamma(p)\omega_c(p)=a\gamma(p_0)\omega_c(p_0)$ and
$\omega_c(p)=\omega_c(p_0)$ \cite{OMS}.

Since, as the quantities $\omega_R$ and $\omega_0$ can slightly
differ and one needs to determine a systematical error caused by
the nonzero difference $\omega_0-\omega_R$, it is necessary to use
general formulae
% Therefore, it is
%natural to consider the case when $<(\omega_0-\omega)^2>\sim
%1/\tau_{coh}^2$, where $\tau_{coh}$ is the spin coherence time.
%The dispersion of
%$\omega_0$ defines the spin coherence time
%$\tau_{coh}=\left<(\delta\omega_0)^2\right>^{-1/2}$. The average
%value of $(\omega_0-\omega)^2$ cannot be less than
%$1/\tau_{coh}^2$.
%It is expedient if the measurement time is of
%order of $\tau_{coh}$ \cite{YOr,YSNotes}. In this case, the
%condition $\Omega't\ll1$ may not be satisfied and we have to use
%the general formula
specifying the spin dynamics. %This formula can be easily derived because
Hereafter the light velocity, $c$, will be explicitly shown.

The dynamics of the vertical component of polarization vector is the
same in the rotating and lab frames. It is given by
\begin{equation}
\begin{array}{c}
P_z(\Phi)=\frac{{\epsilon'}}{\nu'}P_0\left\{\sin{(\psi-\zeta')}\sin{(\nu'\Phi)}
%\right.\\\left.
+\frac{a\gamma_0\!-\!\nu_R}{\nu'}
\cos{(\psi-\zeta')}\left[1-\cos{(\nu'\Phi)}\right]\right\},\\
\epsilon'=\frac{e\eta}{8\pi c
p_0}E_0l\left(\frac{\omega_0}{a\gamma_0^2\omega}\pm1\right),
\end{array}\label{eqhnu}\end{equation}
where $\epsilon'$ is the amplitude of the resonance strength
($|\epsilon'|=|\epsilon^\pm_R|$) and
$P_0$ is the %degree of
initial beam polarization.

%Evidently, $$\nu'\Phi=\frac{\Omega'\Phi}{|\omega_c|}.$$

% It is important that the substitution $\nu'\rightarrow-\nu'$ does not changes the values
%of the component of polarization vector and therefore $\nu'$ and $\nu'\Phi$ can be
%replaced by $\Omega'/\omega_c$ and $\Omega't$, respectively.

Eqs. (\ref{eqrf}),(\ref{eqt}) define the evolution of other
components of polarization vector:
\begin{equation}
\begin{array}{c}
P_\rho(\Phi)=P_0\biggl\{\cos{(\nu'\Phi)}\cos{(\nu_R\Phi+\psi)}+\frac{{\epsilon'}^2}{{\nu'}^2}\cos{(\psi-\zeta')}\left[1-\cos{(\nu'\Phi)}\right]\cos{(\nu_R\Phi
+\zeta')}\\-\frac{a\gamma_0\!-\!\nu_R}{\nu'}\sin{(\nu'\Phi)}\sin{(\nu_R\Phi
+\psi)}\biggr\}, \\
P_\phi(\Phi)=P_0\biggl\{\cos{(\nu'\Phi)}\sin{(\nu_R\Phi+\psi)}+\frac{{\epsilon'}^2}
{{\nu'}^2}\cos{(\psi-\zeta')}\left[1-\cos{(\nu'\Phi)}\right]\sin{(\nu_R\Phi
+\zeta')}\\+\frac{a\gamma_0\!-\!\nu_R}{\nu'}\sin{(\nu'\Phi)}\cos{(\nu_R\Phi+\psi)}\biggr\}.
\end{array}\label{eqhn}\end{equation}

It is important that the substitution $\nu'\rightarrow-\nu'$ does
not change the values of the components of polarization vector and
therefore the quantities $\nu'$ and $\nu'\Phi$ can be replaced by
$\Omega'/\omega_c$ and $\Omega't$, respectively.

When $\nu'\Phi\ll1~(\Omega't\ll1)$,
\begin{equation}\begin{array}{c}
P_\rho(\Phi)=P_0\left[\cos{(\nu_R\Phi+\psi)}-(a\gamma_0\!-\!\nu_R)\Phi\sin{(\nu_R\Phi+\psi)}\right],\\
P_\phi(\Phi)=P_0\left[\sin{(\nu_R\Phi+\psi)}+(a\gamma_0\!-\!\nu_R)\Phi\cos{(\nu_R\Phi+\psi)}\right],
\end{array}\label{eqlle}\end{equation}
\begin{equation}\begin{array}{c}
P_z=P_0\epsilon'\Phi\sin{(\psi-\zeta')}
\\=
\frac{e\eta}{8\pi c
p_0}P_0E_0l\left(\frac{\omega_0}{a\gamma_0^2\omega}\pm1\right)\omega_ct\sin{(\psi-\zeta')}.
\end{array}\label{eqfa}
\end{equation}

For the deuteron EDM experiment,
$\omega\approx\omega_0,~\zeta'=\varphi$, and the plus sign should
be chosen in Eq. (\ref{eqfa}). When the angle
$\Upsilon=\psi-\pi/2$ characterizing an initial spin direction
about the $\bm e_\phi$ axis is used,
$$\sin{(\psi-\zeta')}=\cos{(\Upsilon-\zeta')}=\cos{(\Upsilon-\varphi)}$$
and Eq. (\ref{eqfa}) takes the form
\begin{equation}\begin{array}{c}
P_z=\frac{e\eta}{8\pi c
p_0}P_0E_0l\left(\frac{1}{a\gamma_0^2}+1\right)\omega_ct\cos{(\Upsilon-\varphi)}\\
=\frac{e\eta}{8\pi\beta_0 c^2
p_0}P_0E_0l\left(\frac{1}{a\gamma_0^2}+1\right)\omega_cL_b\cos{(\Upsilon-\varphi)}.
\end{array}\label{eqfap}
\end{equation}
Eqs. (\ref{velma}),(\ref{eqfap}) result in
\begin{equation}\begin{array}{c}
P_z=-\frac{1}{4}\eta P_{0}\Delta\beta_m\gamma_0\left(1+
a\gamma_0^2\right)\omega_ct\cos{(\Upsilon-\varphi)}.
\end{array}\label{eqfnf}
\end{equation}

The corresponding formula obtained in Refs. \cite{YOr,OMS} is
given by
\begin{equation}\begin{array}{c}
P_z=\frac{1}{4}\eta P_{0}\Delta\beta_m\gamma_0\omega_ct,
\end{array}\label{eqYOr}
\end{equation}
where the designations accepted in the present work are used. Only
the special case of $\omega_0=\omega,~\Psi=\varphi$ has been
considered in Refs. \cite{YOr,OMS}.

Eq. (\ref{eqYOr}) differs from Eq. (\ref{eqfnf}) by the absence of
the factor $(1+a\gamma_0^2)$. It is quite natural because the
effect of electric field on the spin dynamics has not been taken
into account in Refs. \cite{YOr,OMS}. The formula obtained in
Refs. \cite{YOr,OMS} should be added by the above mentioned
factor. The signs in Eqs. (\ref{eqfnf}) and (\ref{eqYOr}) are
opposite because the quantity $\omega_c$ was supposed to be
positive in Refs. \cite{YOr,OMS}.

\section {Contributions from Electric and Magnetic Fields in the Experiment
Based on the Frozen Spin Method}

It is of interest to compare ratio (\ref{eqr}) with the
corresponding ratio for the EDM experiment based on the frozen
spin method.

The radial electric field in %the deuteron EDM experiment based on the frozen spin method
%latter
this experiment is adjusted to \cite{EDM}
$$ E=\frac{a\beta\gamma^2}{1-a\beta^2\gamma^2}B. $$
The ratio of contributions from the electric and magnetic fields
to the BVP caused by the EDM is
\begin{equation} \kappa=\frac{E}{|\bm\beta\times\bm
B|}=\frac{a\gamma^2}{1-a\beta^2\gamma^2}.
\label{eq1}\end{equation}

For the deuteron experiment \cite{EDMP},
$a=a_d=-0.14299,~\beta=0.35,~\gamma=1.068,$ and $\kappa=0.17$.
Therefore, the effect of the electric field on the BVP cannot be
neglected in the deuteron EDM experiment based on the frozen spin
method. The contribution from the electric field is negligible in
a similar muon experiment ($a_\mu=1.1659\times 10^{-3}$) and
predominant in a proton one ($a_p=1.7928$). %$a_p=1.7928473$

%We can fulfil corresponding calculations for the resonant EDM
%experiment.

\section {Summary}

The BVP in the resonant EDM experiment is affected by the electric
and magnetic fields. The effect of the resonant electric field is
significant, while the contribution from the magnetic field
proportional to the oscillating part of particle velocity is
dominant. The effective fields defining the resonant effect have
been expressed in terms of the resonance strengths. The spin
dynamics in the resonant deuteron EDM experiment has been
calculated in the general case. In previous works, only a special
case has been considered and the effect of the electric field on
the spin dynamics has not been taken into account. The
electric-field correction is
important because it leads to decreasing the EDM effect by %about
23 percent for the deuteron experiment and increasing it by 12
percent for the proton one.

\section* {Acknowledgements}

The author would like to thank Y.K. Semertzidis for helpful
discussions and acknowledge a financial support by BRFFR (grant
No. $\Phi$06P-074).

%\footnotesize
%\renewcommand{\baselinestretch}{1.4}

\end{document}